# Effects of Zn substitution in $La_{2-x}Sr_xCu_{1-y}Zn_yO_4$: interplay among superconductivity, pseudogap, and stripe order

**S. H. Naqib\* and R. S. Islam**

Department of Physics, University of Rajshahi, Raj-6205, Bangladesh

\*E-mail: salehnaqib@yahoo.com

**Abstract**. The effect of Zn substitution on the superconducting transition temperature, $T_c$, was investigated for the $La_{2-x}Sr_xCu_{1-y}Zn_yO_4$ (Zn-LSCO) compounds over a wide range of hole concentration, $p$ ($\equiv$ x), and Zn content ($y$) in the $CuO_2$ plane. Zn induced rate of suppression of $T_c$, $dT_c(p)/dy$, was found to be strongly $p$-dependent and showed a systematic variation with hole concentration, except in the vicinity of $p \sim 0.12$, *i.e.*, near the so-called $1/8^{th}$ anomaly where the charge/spin stripe correlations are at its strongest in $La_{2-x}Sr_xCuO_4$. Near $p \sim 0.12$, the static striped charge ordering is widely believed to dominate the *T-p* phase diagram. $dT_c(p)/dy$ decreased strongly around this hole concentration *i.e.*, Zn became less effective in degrading $T_c$ near $p \sim 0.12$. This observation is indicative of an intricate and competing interplay among the superconducting, pseudogap, and stripe correlations in Zn-LSCO.

## 1. Introduction

High-$T_c$ cuprate superconductors are remarkable materials as various strongly correlated electronic ground states (*e.g.*, Mott-Hubbard insulating antiferromagnet, spin-glass, pseudogap region, spin/charge ordering) are uncovered with increasing the doped hole concentration, $p$, in the $CuO_2$ planes and a nearly parabolic $T_c(p)$ dome is realized in the doping range $0.05 < p < 0.27$ for most of the cuprate families. Over most of the regions of the *T-p phase diagram* the normal state charge transport and magnetic properties are anomalous in the sense that canonical Fermi-liquid like behavior is only observed in the deeply overdoped (OD) region [1 – 3]. Besides superconductivity itself, pseudogap and stripe correlations are probably the two of the most widely investigated phenomena. The pseudogap (PG) correlation is detected in the *T-p phase diagram* over a certain doping range, extending from the underdoped (UD) to the slightly overdoped regions. In the PG region a number of anomalies are observed both in normal and superconducting states, where contrary to one of the central tenants of the Fermi liquid theory, low-energy excitations are gapped along certain directions of the Brillouin zone while Fermi arcs exist in other directions [3, 4]. It is widely believed that understanding the physics of PG is one of the outstanding obstacles in the path of unlocking the mystery of cuprate superconductivity. Existing scenarios to explain the origin and the *p*-dependence of the PG could be classified into two groups [1, 5]. One is based on the precursor pairing scenario, where PG arises from strong fluctuations of superconducting (SC) origin in the strong coupling regime for systems with low dimensionality (high structural and physical anisotropy) and low superfluid density [6]. In the other scenario, PG is attributed to some correlations of non-SC origin. In this scenario PG coexists and, in fact, often competes with superconductivity [1]. Considerable debate has ensued as to the nature of the PG and no consensus has been reached yet [1, 5, 6]. On the other hand the static spin/charge stripe correlations are only observed in the UD cuprates in the vicinity of $p \sim 0.12$ (the so-called $1/8^{th}$ anomaly) [7, 8], although dynamical (fluctuating) stripe correlations are believed to exist over a much wider doping range, specially in the La214 compounds [9]. Stripe phase occurs as a compromise between the AFM correlations among the Cu spins and Coulomb interaction between the electrons (both favoring localization) and the kinetic energy of the doped charge carriers (leading to delocalization). Broadly speaking, stripe phase can be viewed as spontaneously separated ordered states of charge-rich (high kinetic energy) and charge-poor (strongly antiferromagnetically correlated) regions throughout the compound. It is more or less agreed that static stripe order is detrimental to superconductivity [9], but the possible influence of dynamical stripes on superconductivity is debatable [9 – 11]. Some of the existing theoretical models link the origins of both superconductivity and PG to the stripe correlations [9 – 12]. It is extremely important to clarify the interplay among these different correlations in order to develop a coherent picture describing the physics of high-$T_c$ cuprates. We have used non-magnetic Zn substitution for planar Cu atoms to investigate these for $La_{2-x}Sr_xCu_{1-y}Zn_yO_4$. A number of previous studies have found Zn-induced pinning (slowing down of fluctuations) of the stripe order [13 – 15] in cuprates. It was also found that Zn substitution does not affect the PG energy scale [16 – 18]. On the other hand non-magnetic Zn suppresses SC most effectively [17 – 19]. In this study we have found that the Zn-induced rate of suppression of $T_c$, $dT_c/dy$, is highly *p*-dependent as found in previous studies [20 – 22]. We have also found that $dT_c/dy$ is greatly reduced in the vicinity of $p \sim 0.12$ where charge/spin ordering is at their strongest. This observation, in our knowledge, has not been reported before. The fact that Zn becomes less effective in degrading $T_c$ near the $1/8^{th}$ anomaly is surprising at the first sight, because at this hole concentration superconductivity is already severely weakened due to the presence of the static stripe order. We discuss the possible implications of these findings in section 3.

## 2. Experimental samples and results

A large number of polycrystalline sintered single-phase samples of $La_{2-x}Sr_xCu_{1-y}Zn_yO_4$ were synthesized by solid-state reaction method using high-purity (> 99.99%) powders. In this paper we have used samples with the following compositions: x = 0.08 (Zn-free only), 0.09, 0.10, 0.11, 0.12, 0.14, 0.15, 0.17 (Zn-free only), 0.19, 0.22, 0.27 (Zn-free only) and y = 0.0, 0.005, 0.01, 0.015, 0.02, 0.024. Samples were characterized by X-ray diffraction (XRD), room-temperature thermopower (*S[290K]*), and low-field (*H* = 1 Oe, *f* = 333.33 Hz) AC susceptibility (ACS) measurements. The details of sample preparation and characterization can be found elsewhere [23]. XRD was used to check the phase-purity, *S[290K]* gave an independent check for the Sr content [24, 25], ACS gave $T_c$ values. The transition widths also provided with information about sample homogeneity. Most of the samples used in this study exhibited sharp SC transitions. Transition widths increase somewhat for the heavily Zn substituted compounds. The Sr contents and therefore, the hole contents reported here were found to be accurate within ± 0.005. $T_c$ was determined as follows: a straight line was drawn at the steepest part of the diamagnetic ACS curve and another one was drawn as the *T*-independent base line associated with negligibly small normal state signal. The intercept of the two lines gave $T_c$ (shown in Fig. 1). *$T_c(p, y)$* values obtained from ACS data agree quite well, where available, with those found in previous studies [26, 27]. Iso-valent Zn substitution for in-plane Cu does not change the hole concentration significantly [16, 17, 25]. $T_c$ was found to decrease almost linearly with increasing Zn concentration. Fig. 2 shows the *$T_c(y)$* data for different values of hole contents. *$dT_c(p)/dy$* were calculated from the slopes of the linear fits to the *$T_c(y)$* data for different fixed values of x. Fig. 2 shows clearly the strongly *p*-dependent nature of *$dT_c(p)/dy$*. The magnitude of *$dT_c(p)/dy$,* except near *p* ~ 0.12, decreases sharply with increasing *p* in the UD to optimally doped region, passing through a minimum in the OD near *p* ~ 0.20 and increases slowly again for further overdoping. Similar behavior was observed for Y123 [22] and Bi2212 compounds [20]. *$dT_c(p)/dy$* in the vicinity of p ~ 0.12 is anomalous, a sharp decrease in the magnitude of *$dT_c(p)/dy$* is found here. This implies that near the $1/8^{th}$ doping Zn becomes significantly less effective in decreasing $T_c$. This effect is seen in Fig. 3.

## 3. Discussion and conclusions

The effect of Zn on $T_c$ has been studied extensively since the early days of cuprate superconductivity. The strongly *p*-dependent rate of suppression of $T_c$ for the various cuprate families can, in general, be explained assuming strong potential (unitary) scattering by non-magnetic Zn in the presence of a $dx^2$-$y^2$ order parameter, at least up to *p* ~ 0.20 [20 – 22]. Evidence favoring such strong pair-breaking scattering is found from the STM study [28]. In this picture, the strong *p*-dependent rate of suppression of $T_c$ due to Zn arises mainly from the *p*-dependent PG energy scale [20 – 22]. The minimum in the magnitude of *dT$_c$(p)/dy* at *p* ~ 0.19 follows naturally as PG vanishes at this hole content [1, 17, 18, 20 – 22]. The gradual increase in the magnitude of *dT$_c$(p)/dy* above *p* ~ 0.19, is indicative of a change in the electronic ground state of different origin [22]. It is important to note that, this scenario supports a non-SC origin for the PG. Here PG competes with superconductivity, at least in the sense that its presence removes QP states which otherwise would have been available for the SC condensate (enhancing the superfluid density). The fact that Zn degrades $T_c$ most effectively but does not affect PG energy scale [16 – 18] also lends further support for the non-SC origin of the PG. The anomaly in *dT$_c$(p)/dy* near p ~ 0.12 has a different physical origin. Earlier studies showed that PG is not affected significantly by the presence of static (or quasi-static) stripe order, a nearly linear decrease in the PG energy scale is found for all the cuprates with increasing *p* with no noticeable feature near the 1/8$^{th}$ anomaly [1, 5, 17, 18]. This is suggests that the stripe and the PG correlations are perhaps not directly related.

The size of the anomaly in *dT$_c$(p)/dy* in the vicinity of *p* ~ 0.12 indicates that the effect of Zn on $T_c$ is dominated by the stripe-ordering in this region. As stated earlier, Zn is believed to pin the fluctuating stripe order [13 – 15]. The pinning mechanism can be attributed to Zn induced enhancement of the AFM correlations, carrier localization, or to the increase in the stripe inertia. Irrespective of the precise mechanism, as static charge/spin ordering competes with and weakens superconductivity, Zn substitution should become more effective in reducing T$_c$ near the 1/8$^{th}$ doping, exactly the opposite effect is found experimentally. The possible reasons can be (i) since spin/charge ordering near *p* ~ 0.12 is already static or quasi-static in the pure compound, Zn substitution plays no significant role in pinning in this region. In this picture Zn substitution in compounds with *p* close to the 1/8$^{th}$ value is "*wasted*" to some extent. This supports the theoretical work by Smith *et al.* [29] describing stripe-pinning as the mechanism for degradation of $T_c$ due to Zn. (ii) A large amount of Zn substitution (up to 2.4% in our samples) destroys the integrity of the static stripe order. Assuming random substitution of Zn in the Cu sites, part of the Zn in the hole-rich region will lead to carrier localization and the other part will replace the antiferromagnetically correlated Cu spins in the hole-poor region, by creating spin vacancies. In such a situation with large number of spin vacancies, a hole from a neighboring domain may hop inside the spin ordered region. This process will weaken the stripe order itself [30]. Here Zn becomes less effective in reducing $T_c$ because the static stripe order is weakened and consequently superconductivity is in fact somewhat enhanced.

In summary, we have reported the effect of Zn on $T_c$ as a function of hole concentration for $La_{2-x}Sr_xCu_{1-y}Zn_yO_4$ over a wide range of compositions. A systematic variation in $dT_c(p)/dy$ was found except near $p \sim 0.12$. We have discussed the possible scenarios for this anomaly close to the $1/8^{th}$ doping. Contrasting effects of Zn on $T_c$, PG energy scale, and stripe correlations indicate that they are unrelated phenomena. A similar conclusion was drawn by Tallon *et al.* [31] from their oxygen isotope exponent measurements for $La_{2-x}Sr_xCu_{1-y}Zn_yO_4$.

**Acknowledgements**

The authors acknowledge Prof. J. R. Cooper, University of Cambridge, UK, and Prof. J. L. Tallon of MacDiarmid Institute and Victoria University of Wellington, New Zealand, for some of the samples used in this study. The authors also thank the IRC in Superconductivity, University of Cambridge, and the MacDiarmid Institute and Industrial Research Ltd., Wellington, for providing with the experimental facilities.

**References**

[1] Tallon J L, Loram J W 2001 *Physica C* **349** 53.

[2] Lee P A 2008 *Rep. Prog. Phys.* **71** 012501.

[3] Damascelli A, Hussain Z, Shen Z -X 2003 *Rev. Mod. Phys.* **75** 473.

[4] Norman M R, Pines D, Kallin C 2005 *Adv. Phys.* **54** 715.

[5] Kresin V Z, Ovchinnikov Y N, Wolf S A 2006 *Phys. Rep.* **431** 231.

[6] Emery V J, Kivelson S A 1995 *Nature* **374** 434.

[7] Tranquada J M, Sternlib B J, Axe J D, Nakamura Y, Uchida S 1995 *Nature* **375** 561.

[8] Abbamonte P, Rusydi A, Smadici S, Gu G D, Sawatzky G A, Feng D L 2005 *Nature Phys.* **1** 155.

[9] Carlson E W, Emery V J, Kivelson S A, Orgad D in *The Physics of Superconductors Vol. II. Superconductivity in Nanostructure, High-Tc and Novel Superconductors, Organic Superconductors,* Eds. Bennemann K H, Ketterson J B, *Springer-Verlag* 2004.

[10] Castellani C, Castro C Di, Grilli M 1995 *Phys. Rev. Lett.* **75** 4650 and Castro C Di, Grilli M, Caprara S, Suppa D 2004 *cond-mat/0408058v1*.

[11] Tranquada J M 2005 *J. Phys.* **IV** *France* 1.

[12] Kivelson S A, Bindloss I P, Fradkin E, Oganesyan V, Tranquada J M, Kapitulnik, Howard C 2003 *Rev. Mod. Phys.* **75** 1201.

[13] Akoshima M, Noji T, Ono Y, Koike Y 1998 *Phys. Rev. B* **57** 7491.

[14] Akoshima M, Koike Y, Watanabe I, Nagamine K 2000 *Phys. Rev. B* **62** 6721.

[15] Risdiana, Adachi T, Oki N, Yairi S, Tanabe Y, Omori K, Koike Y, Suzuki T, Watanabe I, Koda A, Higemoto W 2008 *Phys. Rev. B* **77** 054516.

[16] Alloul H, Mendels P, Casalta H, Marucco J F, Arabski J 1991 *Phys. Rev. Lett.* **67** 3140.

[17] Naqib S H, Cooper J R, Tallon J L, Panagopoulos C 2003 *Physica C* **387** 365.

[18] Naqib S H, Cooper J R, Tallon J L, Islam R S, Chakalov R A 2005 *Phys. Rev. B* **71** 054502.

[19] Tarascon J M, Greene L H, Barboux P, Mckinnon W R, Hull G W, Orlando T P, Delin K A, Foner S, McNiff Jr. E J 1987 *Phys. Rev. B* **36** 8393.

[20] Tallon J L 1998 *Phys. Rev. B* **58** 5956.

[21] Williams G V M, Haines E M, Tallon J L 1998 *Phys. Rev. B* **57** 146.

[22] Naqib S H 2007 *Supercond. Sci. Tech.* **20** 964.

[23] Islam R S 2005 *Ph.D. thesis* (University of Cambridge, UK), (unpublished).

[24] Obertelli S D, Cooper J R, Tallon J L 1992 *Phys. Rev. B* **46** 14928.

[25] Tallon J L, Cooper J R, DeSilva P S I P N, Williams G V M and Loram, J W 1995 *Phys. Rev. Lett.* **75** 4114.

[26] Panagopoulos C, Tallon J L, Rainford B D, Xiang T, Cooper J R, Scott C A 2002 *Phys. Rev. B* **66** 064501.

[27] Kastner M A, Birgeneau R J, Shirane G, Endoh Y 1998 *Rev. Mod. Phys.* **70** 897.

[28] Pan S H, Hudson E W, Lang K M, Eisaki H, Uchida S, Davis J C 2000 *Nature* **403** 746.

[29] Smith C M, Castro Neto A H, Balatsky A V 2000 *cond-mat/0012080*.

[30] Anegawa O, Okajima Y, Tanda S, Yamaya K 2001 *Phys. Rev. B* **63** 140506.

[31] Tallon J L, Islam R S, Storey J, Williams G V M, Cooper J R 2005 *Phys. Rev. Lett.* **94** 237002.

**Figure captions**

Figure 1 (color online): Determination of $T_c$ from the ACS for two $La_{2-x}Sr_xCu_{1-y}Zn_yO_4$ samples (see text for details). Hole and Zn contents are shown.

Figure 2 (color online): $T_c(y)$ for different values of hole concentrations. The straight lines are fits to the $T_c(y)$ data. For clarity some of the compounds with other x-values are not shown here.

Figure 3 (color online): $T_c(y = 0.0)$ and $dT_c/dy$ versus *p*. The thick dotted lines are drawn as guides to the eyes.

Figure 1

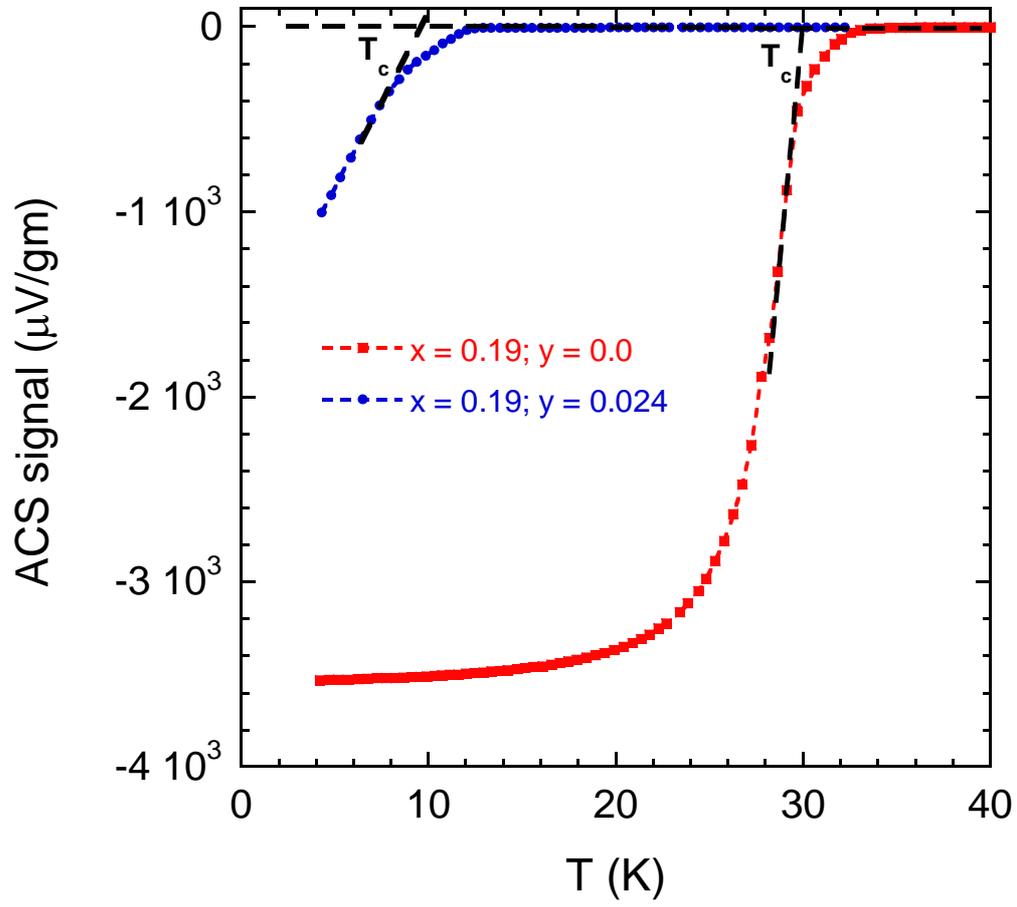

Figure 2

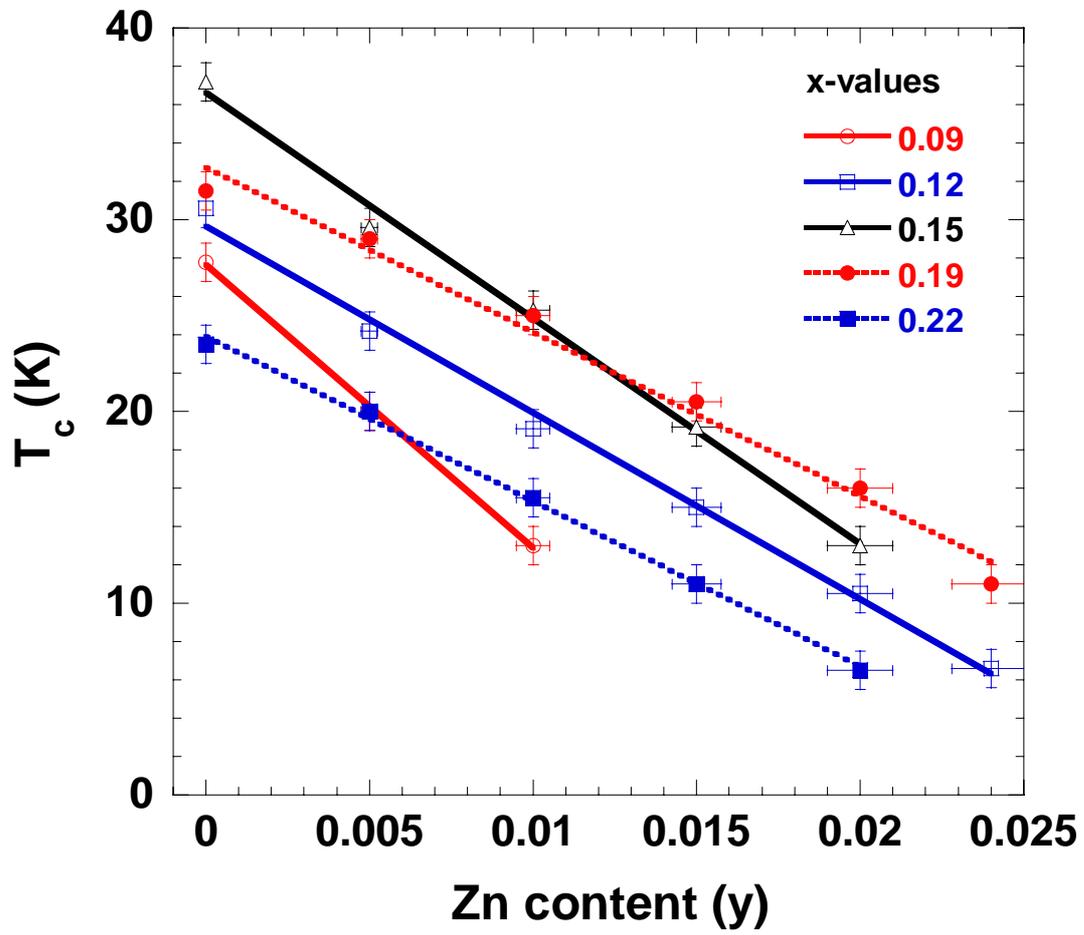

Figure 3

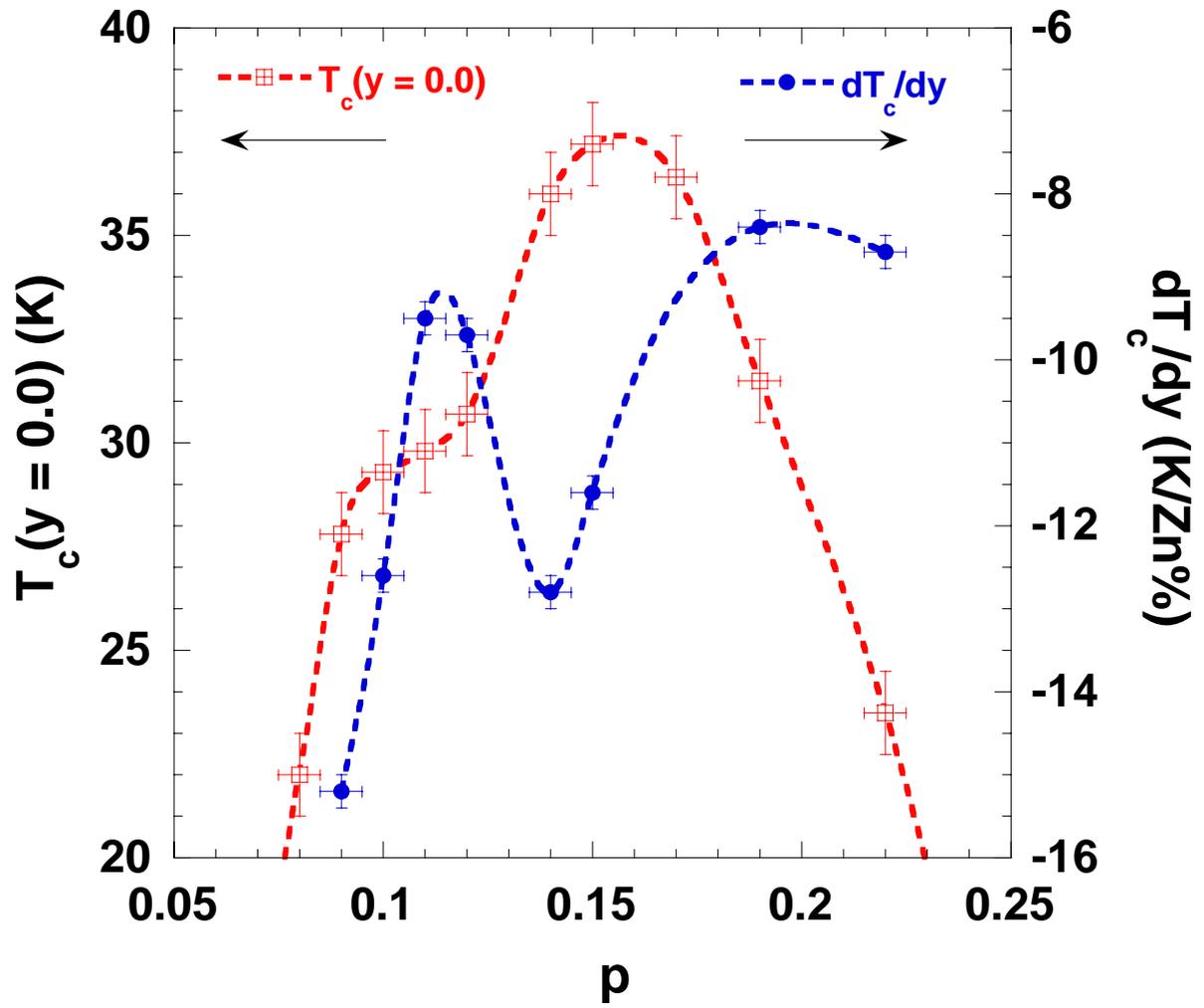